\documentclass[runningheads]{llncs}

\usepackage{graphicx}

\usepackage{paralist}
\usepackage{amssymb,amsmath}
\usepackage{hyperref}
\usepackage{xcolor}
\usepackage{booktabs}
\usepackage{xspace}
\newcommand{\ra}[1]{\renewcommand{\arraystretch}{#1}}
\def\holfour{\textsf{HOL4}\xspace}

\def\cakeml{\textsf{CakeML}\xspace}

\def\tactictoe{\textsf{TacticToe}\xspace}

\def\holyhammer{\textsf{HOL(y)Hammer}\xspace}

\def\opentheory{\textsf{Open Theory}\xspace}

\def\thfone{\textsf{TH1}\xspace}
\def\tffone{\textsf{TF1}\xspace}
\def\tffzero{\textsf{TF0}\xspace}
\def\thfzero{\textsf{TH0}\xspace}
\def\fof{\textsf{FOF}\xspace}
\def\tptp{\textsf{TPTP}\xspace}

\def\num{\mathsf{num}}

\def\limplies{\Rightarrow}
\newcommand{\todo}[1]{}

\makeatletter
\renewcommand\section{\@startsection{section}{1}{\z@}%
                       {-12\p@ \@plus -4\p@ \@minus -4\p@}%
                       {8\p@ \@plus 4\p@ \@minus 4\p@}%
                       {\normalfont\large\bfseries\boldmath
                        \rightskip=\z@ \@plus 8em\pretolerance=10000 }}
\makeatother

\begin{document}

\title{GRUNGE: A Grand Unified ATP Challenge\thanks{Supported by the ERC
    % Consolidator
    grant no. 649043 AI4REASON and no. 714034 SMART, by the Czech project AI\&Reasoning 
CZ.02.1.01/0.0/0.0/15\_003/0000466, the European Regional Development Fund,
and the National Science Foundation Grant 1730419 - ``CI-SUSTAIN: StarExec: 
Cross-Community Infrastructure for Logic Solving''.}}
\author{Chad E. Brown\inst{1}
\and   Thibault Gauthier\inst{1}
\and Cezary Kaliszyk\inst{2}
\and \\Geoff Sutcliffe\inst{3}
\and Josef Urban\inst{1}}
\authorrunning{Brown et al.}
\institute{Czech Technical University in Prague
  \and
  University of Innsbruck and University of Warsaw
  \and
  University of Miami}
\maketitle            
\begin{abstract}
  This paper describes a large set of related theorem proving problems
  obtained by translating theorems from the HOL4 standard library into 
  multiple logical formalisms.
  The formalisms are in higher-order logic (with and without type variables)
  and first-order logic (possibly with types, and possibly with 
  type variables).
  The resultant problem sets allow us to run automated theorem provers that 
  support different logical formalisms on corresponding problems, and compare 
  their performances.
  This also results in a new ``grand unified'' large theory benchmark
  that emulates the ITP/ATP hammer setting, where systems and metasystems
  can use multiple formalisms in complementary ways, and jointly
  learn from the accumulated knowledge.
\end{abstract}

\section{Introduction}\label{sect:introduction}

A hammer~\cite{hammers4qed}  for an interactive theorem prover 
(ITP)~\cite{HarrisonUW14} typically translates an ITP goal into a formalism 
used by an automated theorem prover (ATP).
Since the most successful ATPs have so far been for untyped first-order logic, 
the focus has been on first-order translations. 
There is also interest in ATPs working in richer formalisms, such as 
monomorphic and polymorphic, typed first-order, and higher-order logics.
The TPTP formats for various formalisms have been adopted for this work, 
viz. \fof~\cite{Sutcliffe17} \tffzero~\cite{SS+12}, \tffone~\cite{tff1}, 
\thfzero~\cite{C25}, and \thfone~\cite{KSR16}.
An interesting related task is the creation of a (grand) unified large-theory 
benchmark that allows fair comparison of such ATP systems, their 
combination and integration with premise selectors and machine 
learners~\cite{abs-1108-3446}, across different formalisms.
As a step towards creating such benchmarks we present two families of
translations from the language of \holfour~\cite{SlindN08} to the various 
TPTP formats.
We have implemented these translations and plan to use them as the first
``GRand UNified ATP challenGE'' (GRUNGE) benchmarks, generalizing existing 
benchmarks such as the \cakeml export~\cite{KumarMNO14} that was used in the 
large-theory benchmark (LTB) division of the CASC-J9 ATP 
competition~\cite{Sutcliffe18}. 

The rest of the paper is structured as follows.
Section~\ref{Preliminaries} introduces notation and the HOL syntax. 
Section~\ref{HOL4} introduces the problems -- the \holfour
standard library. 
Section~\ref{thibault} introduces the first family of translations, and
Section~\ref{chad} introduces the second family of translations.
Section~\ref{case} discusses and compares the translations on an example,
and Section~\ref{results} evaluates the translations using existing ATPs.
Section~\ref{CASCLTB} describes the CASC-27 LTB division, which is based
on these translations.
Related work is discussed in Section~\ref{relatedwork}.

\section{Preliminaries}
\label{Preliminaries}
    
Since this work is based on the \holfour standard library, it is necessary
to start with brief comments about the syntax and notion of proof in \holfour.
More detailed information is in~\cite{SlindN08,Gordon:1993:ITH}.
\holfour, like several other ITPs (e.g., Isabelle/HOL~\cite{NPW02}, 
HOL Light~\cite{Harrison96} or ProofPower~\cite{king1996development}), 
is based on an extension of Church's simple type 
theory~\cite{Church:SimplyTyped} that includes
prefix polymorphism and type definitions~\cite{Gordon:1993:ITH}.
\holfour includes a type $o$ of propositions, a type $\iota$ of individuals
and a type $(\sigma\to \tau)$ of functions from a type $\sigma$ to a 
type $\tau$.
Parentheses are omitted, with $\to$ associating to the right.
In addition, there are type variables $\alpha$ and defined types.
At each point in the development of the \holfour library there is a finite 
set of (previously defined) constants $c$, and a a finite set of
(previously defined) type constructors $\kappa$ giving a type 
$\kappa(\sigma_1,\ldots,\sigma_n)$ for types $\sigma_1,\ldots,\sigma_n$.
For simplicity we consider the signature to be fixed, to avoid the need
to specify the types and terms relative to an evolving signature.

Terms are generated from constants $c$ and variables $x$ using application 
$(s\ t)$ and $\lambda$-abstractions $(\lambda x:\sigma.t)$ in the expected way,
for terms $s$ and $t$. 
Parentheses are omitted, with application associating the left. 
Binders have scope as far to the right as possible, consistent with 
parentheses. 
Multiple binders over the same type can be written in a combined form, e.g.,
$\lambda x y:\sigma.t$ means $\lambda x:\sigma.\lambda y:\sigma.t$.

Constants may be polymorphic.
There are two primitive polymorphic logical constants:
$=^\alpha$ is polymorphic with type $\alpha\to\alpha\to o$
and $\varepsilon^\alpha$ is polymorphic with type $(\alpha\to o)\to\alpha$,
where $\alpha$ is a type variable.
When terms are defined, such constants are used with a fixed type for $\alpha$
written as a superscript.
New polymorphic constants can be defined within a \holfour theory.

Aside from $=^\alpha$ and $\varepsilon^\alpha$, implication $\limplies$ of 
type $o\to o\to o$ is primitive. 
From these primitive logical constants it is possible to define $\land$, 
$\lor$ and $\neg$, as well as polymorphic operators $\forall^\alpha$ and 
$\exists^\alpha$.
The usual notation is used for these logical connectives, so that the
binder notation $\forall^\sigma (\lambda x:\sigma.t)$ is written as 
$\forall x:\sigma.t$, using the same binding conventions as for 
$\lambda$-abstractions.
Similarly, $\exists^\sigma (\lambda x:\sigma.t)$ is written as 
$\exists x:\sigma.t$.

Terms of type $o$ are called propositions, and we use $\varphi$ and $\psi$
to range over propositions.
A sequent is a pair $\Gamma\vdash \varphi$ where $\Gamma$ is a finite set 
of propositions and $\varphi$ is a proposition.
There is a notion of \holfour provability for sequents.
While our translations map \holfour sequents to TPTP formulae, it is not
our intention to mirror \holfour provability in the target format.
The intention, roughly speaking, is to gain information about when a 
\holfour theorem is a consequence of previous \holfour theorems, in some logic 
weaker than \holfour.

From the simplest perspective, each translation translates \holfour types and 
\holfour terms (including propositions) to terms in the target format.
A type, term or sequent with no type variables is called {\emph{monomorphic}}.
As an optimization, some of the translations translate some monomorphic 
\holfour types to types in the target language.
As a common notation throughout this paper, a \holfour type $\sigma$ 
translated as a term is written as $\hat{\sigma}$, and $\sigma$ translated
as a type is written as $\tilde{\sigma}$.
Another optimization is to translate \holfour propositions (and sequents)
to the level of formulae in the target language.

\section{Problem Set: The \holfour Standard Library}
\label{HOL4}
Version Kananaskis-12 of the HOL4 standard library
contains 15733 formulae: 8 axioms, 2294 definitions, and 13431 theorems.
If most of the formulae were monomorphic and fell into a natural first-order 
fragment of \holfour, then there would be a natural translation into the
\fof format.
However, many formulae are either polymorphic or higher-order (or both), as 
Table~\ref{tab:stats} shows (note that the numbers are not cumulative, e.g.,
the 2232 monomorphic first-order formulae do not include the 1632 uni-sorted
first-order formulae, which could also be processed by an ATP that can handle
the monomorphic types).
The problem set consists of 12140 theorems proven in the \holfour standard 
library\footnote{1291 theorems were not included due to dependencies being
erased during the build of the \holfour library.},
in the context of a finite set of dependencies used in the 
\holfour proof~\cite{hh4h4}.

\begin{table}[htb]
\begin{center}
\begin{tabular}{lrlrlrl}
\toprule
 & \multicolumn{2}{c}{First-order} & \multicolumn{2}{c}{Higher-order} & 
 \multicolumn{1}{c}{Combined} \\
 \midrule
 Uni-sorted  & 1632 & (\fof)     &     0 &            & 1632 \\
 Monomorphic & 2232 & (\tffzero) &  3683 & (\thfzero) & 5915 \\
 Polymorphic & 1536 & (\tffone)  &  6650 & (\thfone)  & 8186 \\
 Combined    & 5400 &            & 10333 &            &15733 \\
 \bottomrule
\end{tabular}
\end{center}
 \caption{\label{tab:stats}Number of \holfour formulae in each category.}
\vspace{-4mm}
\end{table}

\section{Syntactic Translations via First-Order Encodings}
\label{thibault}

There already exist many families of translations for HOL to the TPTP format,
usually developed for hammers~\cite{holyhammer,MengP08}.
We have adopted and adapted them:
(i)~We have made the translations more local, by making them independent for 
each theorem, i.e., unaffected by other theorems. In particular,
this means that when the same lambda function appears in two theorems,
lambda-lifting will produce two new functions.
(ii)~We have made more problems provable (in principle) by introducing 
additional axioms and relying on an embedding of polymorphic types instead 
of relying on heuristic monomorphization.
(iii)~For the \tffzero and \thfzero formats, we have made use of their 
polysortedness by expressing the type of monomorphic constants directly 
using \tffzero and \thfzero types.

The translations are described in the order $\thfone \rightarrow \tffone 
\rightarrow \fof \rightarrow \tffzero \rightarrow \thfzero$, as translations 
to the later formats take advantage of translation techniques used for earlier 
formats.

\subsection{Translating to \thfone}
\label{sec:tg_th1}

\thfone is a language that is strictly more expressive than \holfour. 
Therefore \holfour formulae can be represented in \thfone with minimal effort.
This produces the \thfone-I collection of ATP problems.

\paragraph {Alignment of Logical Constructions.}
The \tptp format contains a set of defined constructs that have implicit 
definitions, and \holfour objects are mapped to their \tptp counterparts
in a natural way. 
The boolean type $o$ of \holfour is mapped to the defined TPTP type {\tt \$o}. 
The arrow type operator is mapped to the TPTP arrow {\tt >}. 
All other type operators are declared to take $n$ types and give a type,
using the TPTP ``type of types'' {\tt \$tType}.
For example, the type operator $\mathit{list}$ has type 
{\tt \$tType > \$tType}.

The TPTP
logical connectives $\wedge$, $\vee$, $\Rightarrow$, $\neg$, $=$, 
$\forall$, $\exists$ are used at the top-level of the translated formula 
whenever possible,
but the corresponding \holfour constants are used when necessary.
Equivalences relating \holfour logical constants to \tptp connectives 
are included.

\paragraph {Explicit type arguments.}
A \holfour constant $c$ carries a type $\nu'$. 
This type is an instance of type $\nu$ that was given to $c$ when $c$ was 
created. 
By matching $\nu$ with $\nu'$, a type substitution $s$ can be inferred. 
Ordering the type variables in the domain of $s$, the implicit type arguments 
of $c$ are deduced. 
Making the quantification of type variables and the type arguments explicit 
is required in the \thfone format. 
The effect that this requirement has on a constant declaration and a 
formula is shown in Example~\ref{tab:explicit_type}.

\begin{example}\label{tab:explicit_type}Explicit type arguments and 
type quantifications

\vspace*{0.5em}
\centering\ra{1.3}
\begin{tabular}{lcc}
 \toprule
 & \holfour & \thfone  \\
 \midrule
Type of $I$ & $\alpha \to \alpha$ & $\forall \alpha:\$tType.\ \alpha 
\rightarrow \alpha$\\
Formula & $\forall x:\alpha.\ (I\ x) = x$ & $\ \ \ \forall \alpha:\$tType.\ 
\forall 
x:\alpha.\ ((I\ \alpha)\ x) =\ x$\\
 \bottomrule
\end{tabular}
\end{example}

\subsection{Translating to \tffone}
\label{tg_tf1}

To produce the \tffone-I collection of ATP problems, all the higher-order 
features of the \holfour problems have to be eliminated.
This is done in a sequence of steps.
\vspace{-3mm}
\paragraph{Lambda-lifting and Boolean-lifting.}
One of the higher-order features is the presence of lambda-abstraction.
The translation first uses the extensionality property to add extra arguments 
to lambdas appearing on either side of an equality, and then beta-reduce the 
formula.
Next lambda-lifting~\cite{DBLP:conf/cpp/Czajka16,MengP08} is used.
Lambda-lifting creates a constant $\mathbf{f}$ for the leftmost 
outermost lambda-abstractions appearing at the term-level. 
This constant $\mathbf{f}$ replaces the lambda-abstraction in the formula.
A definition is given for $\mathbf{f}$, which may involve some variable 
capture - see Example~\ref{ex:lambda-lifting}.
This procedure is repeated until all the atoms are free of lambda-abstractions. 
The definitions are part of the background theory, and are considered to
be axioms in the \tffone problem even if they were created from the conjecture.

\begin{example}\label{ex:lambda-lifting}Lambda-lifting with variable capture
\begin{align*}
\mbox{Original formula:} \ \ & \forall k.\ \mathsf{linear}\ (\lambda x.\ k 
\times x)\\
\mbox{Additional definition:} \ \ &\forall k\ x.\ \mathbf{f}\ k\ x = k \times 
x\\
\mbox{New formula:} \ \ &\forall k.\ \mathsf{linear}\ (\mathbf{f}\ k)
\end{align*}
\end{example}

A similar method can be applied to move some logical constants from the 
term-level to the formula-level - see Example~\ref{ex:boolean-lifting}
(this optimization is not applied in our second family of translations).

\begin{example}\label{ex:boolean-lifting}Boolean-lifting
\begin{align*}
\mbox{Original formula:} \ \ &\forall x.\ x = \mathsf{COND}\ (x=0)\ 0\ x\\
\mbox{Additional definition:} \ \ &\forall x.\ \mathbf{f}\ x \Leftrightarrow (x 
= 0)\\
\mbox{New formula:} \ \ &
\forall x.\ x = \mathsf{COND}\ (\mathbf{f}\ x)\ 0\ x
\end{align*}
\end{example}

To allow the ATPs to create their own encoded lambda-abstractions, axioms 
for the combinators $\mathsf{S}$, $\mathsf{K}$ and $\mathsf{I}$ are added
to every problem.
These axioms are omitted in the second family of translations.
In the \thfzero-II versions of the
problem, combinators are not needed since all simply typed $\lambda$-calculus
terms are already representable. In the \tffzero-II and \fof-II versions only
an axiom for $\mathsf{I}$ and a partially applied axiom for $\mathsf{K}$
are included. Combinator axioms enlarge the search space, which  
hinders the ATPs unnecessarily because they are not needed for proving 
most of the theorem. 

\vspace{-2mm}
\paragraph{Apply operator and arity equations.}
As functions cannot be passed as arguments in first-order logic, an explicit 
apply operator $\mathsf{ap}$ is used to apply a function to an argument.
This way all objects (constants and variables) have arity zero except for 
the apply operator which has arity two.
The \holfour functional extensionality axiom is added to all problems, 
as it expresses the main property of the apply operator:
\[\forall f\ g\ (\forall x.\ \mathsf{ap}\ f\ x = \mathsf{ap}\ g\ x) \Rightarrow 
f = g\]

This axiom also demonstrates how the higher-order variables $f$ and $g$ become
first-order variables after the introduction of $\mathsf{ap}$.

To limit the number of apply operators in a formula, versions of each constant 
are defined for all its possible arities, in terms of the zero-arity version.
These constants are used in the translated formula -
see Example~\ref{ex:arity}.

\begin{example}\label{ex:arity}Using constants with their arity
\begin{align*}
\mbox{Original formula:} \ \ & \mathsf{SUC}\ \mathsf{0} = \mathsf{1} \wedge 
\exists y.\ 
\mathsf{MAP}\  \mathsf{SUC}\ y \neq y\\
\mbox{Arity equations:} \ \ & \mathsf{SUC_1}\ x = \mathsf{ap}\ \mathsf{SUC_0} 
\ x,\ \ldots\\
\mbox{New formula:} \ \ &\mathsf{SUC}_1\ \mathsf{0_0} = \mathsf{1_0} \wedge 
\exists y.\ 
\mathsf{MAP}_2\ 
\mathsf{SUC}_0\ y \neq y
\end{align*}
\end{example}

If the return type of a constant is a type variable then some of its instances 
can expect an arbitrarily large numbers of arguments. 
In the case where the number of arguments $n'$ of an instance exceeds the 
number of arguments $n$ of the primitive constant, variants of this constant 
are not created for this arity.
Instead, the apply operator is used to reduce the number of arguments to $n$.
For example, the term $\mathsf{I}\ \mathsf{I}\ x$ is not translated to 
$\mathsf{I_2}\ \mathsf{I_0}\ x$ but instead to 
$\mathsf{ap}\ (\mathsf{I_1}\ \mathsf{I_0})\ x$.

\vspace{-2mm}
\paragraph{\tffone types.}
As a final step, type arguments and type quantifications are added as in 
Section~\ref{sec:tg_th1}\ . 
Moreover, the boolean type of \holfour is replaced by {\tt \$o} at the 
formula-level, and by $o$ at the term-level (because {\tt \$o} is not allowed 
at the term-level in a first-order formula). 
This causes a mismatch between the type of the atom $o$ and the type of the
logical connective {\tt \$o}. 
Therefore an additional operator $\mathsf{p}\!:\!o \rightarrow \$o$ is applied
on top of every atom. 
The following properties of $\mathsf{p}$ and $o$ are added to every 
translated problem (written here in the first-order style for function 
application):
\begin{align*}
&\forall x y\!:\!o.\ (\mathsf{p}(x) \Leftrightarrow \mathsf{p}(y)) \Rightarrow 
(x=y)\\
&\mathsf{p}(\mathsf{true}),\ \neg \mathsf{p}(\mathsf{false}),\ \forall 
x\!:\!o.\ x = \mathsf{true} \vee x = \mathsf{false}
\end{align*}

In a similar manner, the TPTP arrow type cannot be used whenever a function
appears as an argument. 
Instead the type constructor $\mathsf{fun}$ is used, as illustrated by 
the following constant declaration:
\begin{align*}
\mathsf{ap} &: \forall \alpha:\$tType\ \beta:\$tType.\ 
((\mathsf{fun}(\alpha,\beta) \times 
\alpha) \rightarrow \beta)).
\end{align*}

\subsection{Translating to \fof}

The translation to \fof, which produces the \fof-I collection of ATP problems,
follows exactly the same path as the translation to \tffone except that the 
types are encoded as first-order terms. 
To represent the fact that a first-order term $t$ has type $\nu$, the tagging 
function $\mathsf{s}$, introduced by Hurd~\cite{hurd03}, is used:
every term $t$ of type $\nu$ is replaced by $\mathsf{s}(\hat{\nu},t)$. 
Going from the type $\nu$ to the term $\hat{\nu}$ effectively transforms type 
variables into term variables, and type operators into first-order functions 
and constants. 
Type arguments are unnecessary as the tags contain enough information.
In practice, the $\mathsf{s}$ tagging function prevents terms of different 
types from unifying, and allows instantiation of type variables - see 
Example~\ref{ex:type-inst}.

\begin{example}\label{ex:type-inst}Type instantiation
\begin{center}
\ra{1.3}
\begin{tabular}{ccc}
\toprule
 \tffone &\phantom{aa}& \fof \\
 \midrule
 $\forall x\!:\!\alpha.\ \mathsf{I} (\alpha,x) = x$   && $\forall \hat{\alpha}\ 
 \hat{x}.\ 
\mathsf{s}(\hat{\alpha},\hat{\mathsf{I}}(\mathsf{s}(\hat{\alpha},\hat{x}))) = 
\mathsf{s}(\hat{\alpha},\hat{x})$\\
 $\forall x\!:\!\num.\ \mathsf{I} (\num,x) = x $ && 
$\forall \hat{x}.\ 
\mathsf{s}(\widehat{\num},\hat{\mathsf{I}}(\mathsf{s}(\widehat{\num},\hat{x}))) 
= 
\mathsf{s}(\widehat{\num},\hat{x})$\\
 \bottomrule
\end{tabular}
\end{center}
\end{example}

\subsection{Translating to \tffzero}

An easy way to translate \holfour formulae to \tffzero, which produces the
\tffzero-I collection of ATP problems, is to take the translation to \fof 
and inject it into \tffzero. 

\paragraph{Trivial Injection from \fof to \tffzero.}
The first step is to give types to all the constants and variables appearing 
the the \fof formula. 
A naive implementation would be to give the type $\iota ^ n \rightarrow \iota$ 
to symbols with arity $n$. 
However, since it is known that the first argument comes from the universe
of non-empty types, and the second argument comes from the universe of
untyped terms, an explicit distinction can be made.
The type of $\mathsf{s}$ is defined to be $\delta \times \mu \rightarrow \iota$,
with $\delta$ being the universe of non-empty types, $\mu$ being the 
universe of untyped terms, and $\iota$ being the universe of typed terms.
After this translation a type operator (or type variable) with $m$ arguments 
has type $\delta ^ m \rightarrow \delta$, and a function (or term variable) 
with $n$ arguments has type $\iota ^ n \rightarrow \mu$. 
The type of $\mathsf{p}$ is $\mathsf{p}: \iota \rightarrow {\tt \$o}$.
Declaring the type of all these objects achieves a trivial translation from 
\fof with tags to \tffzero.

\vspace{-2mm}
\paragraph{Using Special Types.}
To take full advantage of the polysortedness of \tffzero, a constant 
$\tilde{c_n}^\nu$ is declared for every constant $c_n$, with arity $n$ and 
monomorphic type $\nu$. 
The type of $\tilde{c_n}^\nu$ is declared to be 
$(\tilde{\nu_1} \times \ldots \times \tilde{\nu_n}) 
\rightarrow \tilde{\nu_0}$, where $\tilde{\nu_1}, \ldots, \tilde{\nu_n}$, and 
$\tilde{\nu_0}$ are basic types. 
A basic type constructs a single type from a monomorphic type, e.g.,
$\mathit{list\_real}$ for $\mathit{list[real]}$, $\mathit{fun\_o\_o}$ for 
$\mathit{fun(o,o)}$.
The basic types are special types, and are declared using {\tt \$tType}.
Thanks to these new constants monomorphic formulae can be expressed in a 
natural way, without type encodings in the formula.
Nevertheless, an ATP should still be able to perform a type instantiation if 
necessary. 
That is why we relate the monomorphic representation with its tagged 
counterpart.

If a term has a basic type then it lives in the monomorphic world 
where as a term of type $\iota$ it belongs to the tagged world.
All monomorphic terms (constructed from monomorphic variables and constants) 
can be expressed in the monomorphic world.
To relate the two representations of the same \holfour term an ``injection'' 
$\mathsf{i}_\nu$ and a ``surjection'' $\mathsf{j}_\nu$ are defined for each 
basic type $\tilde{\nu}$.
The constants $\mathsf{i}_\nu: \tilde{\nu} \rightarrow \mu$ and  
$\mathsf{j}_\nu: \iota \rightarrow \tilde{\nu}$ must respect the 
following properties, which are included as axioms in the translated problems:
\begin{align*}
&\forall x\!:\!\mu.\ 
\mathsf{s}(\hat{\nu},\mathsf{i}_\nu(\mathsf{j}_\nu(\mathsf{s}(\hat{\nu},x)))) = 
\mathsf{s}(\hat{\nu},x)\\
&\forall x\!:\!\tilde{\nu}.\ 
\mathsf{j}_\nu(\mathsf{s}(\hat{\nu},\mathsf{i}_\nu(x))) 
= x
\end{align*}

Whenever $\tilde{c_n}^\nu$ is an instance of a polymorphic function 
$\hat{c_n}$, the following equation is included in the \tffzero problem, 
which relates the two representatives:
\[\forall x_1\!:\!\tilde{\nu_1}\!.\!.\!.x_n\!:\!\tilde{\nu_n}.\ 
\mathsf{s}(\nu_0,\mathsf{i}_{\nu_0}(\tilde{c_n}^\nu(x_1,\!.\!.\!.,x_n)))
 =
\mathsf{s}(\nu_0,\hat{c_n}(
\mathsf{s}(\nu_1,\mathsf{i}_{\nu_1}(x_1)),\!.\!.\!., 
\mathsf{s}(\nu_n,\mathsf{i}_{\nu_n}(x_n)))\]

Example~\ref{ex:fof-tf0} shows how ``injections'', 
``surjections'', and special types can be used to translate a theorem mixing 
polymorphic and monomorphic elements.

\begin{example}~\label{ex:fof-tf0}Special types\\
The polymorphic function $I$ is applied to the monomorphic 
variable $x$. Using special types, type tags can be dropped for $x$.

\vspace{-5mm}

\begin{align*}
\mbox{\fof:\ } &\ \forall \hat{x}.\ 
\mathsf{s}(\widehat{\num},\hat{\mathsf{I}}(\ 
\mathsf{s}(\widehat{\num},\hat{x})\ )) = 
\mathsf{s}(\widehat{\num},\hat{x})\\
\mbox{\tffzero:\ } &\ 
\forall \tilde{x}:\!\widetilde{\num}.\ 
\mathsf{j}_{\num}(\hat{\mathsf{I}}(\ 
\mathsf{s}(\widehat{\num},\mathsf{i}_{\num}(\tilde{x}))\ )) = \tilde{x}\\
\end{align*}
\end{example}

\vspace{-10mm}

\paragraph{Effect on defined operators.}
The $\mathsf{ap}$ operator is treated in the same way as every other constant. 
In particular, a different version of $\mathsf{ap}$ is created for each 
monomorphic type.
The type of $\mathsf{p}$ becomes $\tilde{o} \rightarrow \$o$, and the 
projection $\mathsf{j}_o$ is used to transfer atoms from the tagged world 
to the monomorphic world.

If the presence of the $\mathsf{p}$ predicate and the inclusion of additional 
equations are ignored, our translation of a \holfour first-order monomorphic 
formula using special types to \tffzero is simply the identity transformation.

\subsection{Translating to \thfzero}

Translating from \holfour to \thfzero, which produces the \thfzero-I collection 
of ATP problems, is achieved in a way similar to the translation to \tffzero.
The \holfour formulae are first translated to \fof and then trivially 
injected into \thfzero. 
Special types are used for basic types extracted from monomorphic types. 
The set of higher-order basic types is slightly different from the 
first-order one, where we recursively remove arrow types until a non-arrow 
constructor is found. 
In the higher-order setting a single monomorphic constant $\tilde{c}^\nu$ 
is used to replace all arity versions of $c$:
$\forall f\ x.\ \tilde{\mathsf{ap}}^\nu\ f\ x = f\ x$.
Another benefit of the expressivity of \thfzero is that the basic type 
$\tilde{o}$ can be replaced by $\$o$, and the the predicate $\mathsf{p}$ 
can be omitted.
The effect of the previous steps is illustrated in Example~\ref{ex:tf0-th0}.

\begin{example}~\label{ex:tf0-th0} Translations of $\exists f.\ f\ 0 = 0$\\
  In this example $\tilde{\textsf{ap}}^{\nu}$ has type 
  $(\mathsf{fun\_num\_num}  \times \widetilde{\num}) \to \widetilde{\num}$
  where $\mathsf{fun\_num\_num}$ is the special type corresponding to 
  $\num\to\num$.

\vspace{-7mm}

\begin{align*}
\mbox{\tffzero:\ } &\ \exists \tilde{f}\!:\!\mathsf{fun\_num\_num}.\ 
\tilde{\textsf{ap}}^{\nu}\ (\tilde{f},\tilde{0}^\num) = \tilde{0}^\num\\
\mbox{\thfzero:\ } &\ \exists \tilde{f}\!:\!\widetilde{\num} \to 
\widetilde{\num}.\ 
\tilde{f}\ 
\tilde{0}^\num = \tilde{0}^\num\\
\end{align*}
\end{example}

\vspace{-5mm}

In order to have the same shallowness result for \thfzero as for \tffzero,
it would be necessary to replace monomorphic constants created by the lifting 
procedure by their lambda-abstractions. 
We chose to keep the definitions for the lifted constants, as they allow some 
term-level logical operators to be pushed to the formula level.

\section{Semantic Translations via Set Theory Encodings}
\label{chad}

The second family of translations into \thfzero, \tffzero, and \fof is
semantically motivated~\cite{Pitts93}: we make use of constructors known to be 
definable in 
set theory.
Types and terms are translated to sets, where types must translate to 
non-empty sets.
The translation may optionally use other special types for monomorphic types 
in the \holfour source.
In the \thfzero case the builtin type {\tt \$o} can be used for the \holfour 
type $o$.
In the first-order cases \holfour terms of type $o$ are sometimes translated
to terms, and sometimes to formulae, depending on how the \holfour term is 
used.
In the \tffzero case a separate type ${\tilde{o}}$ of booleans is declared,
which is used as the type of terms translated from \holfour terms of type $o$.
In the \fof case this approach is not possible, as all terms have the same 
type (intuitively representing sets).
The other main difference between the translation to \thfzero and the 
translations to the first-order languages is that the first-order translations 
make use of lambda lifting~\cite{DBLP:conf/cpp/Czajka16,MengP08}.
As a result of the translations we obtain three new collections of ATP 
problems are produced: \thfzero-II, \tffzero-II and \fof-II.

\subsection{Translating to \thfzero}

The base type $o$ for propositions is written as {\tt \$o} in \thfzero,
and $\iota$ for individuals is written as {\tt \$i}.
In addition a base type $\delta$ is declared.
The translation treats elements of type $\iota$ as sets, and elements of type 
$\delta$ as non-empty sets.
The basic constants used in the ATP problems are as follows:
\begin{compactitem}
\item $\mathsf{bool}:\delta$ is used for a fixed two element set.
\item $\mathsf{ind}:\delta$ is used for a fixed non-empty set corresponding 
      to \holfour's type of individuals.
\item $\mathsf{arr}:\delta\to\delta\to\delta$ is used to construct the 
      function space of two sets. 
\item $\mathsf{mem}:\iota\to\delta\to o$ corresponds to the membership 
      relation on sets, where the second set is known to be non-empty.
      The term $\mathsf{mem}~s~t$ is written as $s\in t$, and the term
      $\forall x.x\in s\rightarrow t$ is written as $\forall x\in s.t$.
\item $\mathsf{ap}:\iota\to\iota\to\iota$ corresponds to set theory level 
      application (represented as a set).
\item $\mathsf{lam}:\delta\to(\iota\to\iota)\to\iota$ is used to build set 
      bounded $\lambda$-abstractions as sets.
\item $\mathsf{p}:\iota\to o$ is a predicate that indicates whether or not 
      an element of $\mathsf{bool}$ is true or not.
\item $\mathsf{i}_o:o\to \iota$ is an injection of $o$ into $\iota$, 
      essentially translating false to a set and true to a different set.
\end{compactitem}

\vspace*{0.5em}
\noindent
The basic axioms included in each ATP problem are:
\begin{description}
\item[${\mathbf{Inj}}_o$:] $\forall X:o.{\mathsf{i}}_o X\in \mathsf{bool}$.
\item[${\mathbf{Iso}}^1_o$:] $\forall X:o.{\mathsf{p}}({\mathsf{i}}_o X) = X$.
\item[${\mathbf{Iso}}^2_o$:] $\forall X\in\mathsf{bool}.{\mathsf{i}}_o 
({\mathsf{p}} X) = X$.
\item[${\mathbf{ap_{tp}}}$:] $\forall A B:\delta.\forall f \in 
(\mathsf{arr}~A~B) . \forall x\in A . (\mathsf{ap}~f~x)\in B$.
\item[${\mathbf{lam_{tp}}}$:] $\forall A B:\delta.\forall 
F:\iota\to\iota.(\forall x\in A.F~x\in B)\rightarrow (\mathsf{lam}~A~F)\in 
(\mathsf{arr}~A~B)$.
\item[${\mathbf{FunExt}}$:] $\forall A B:\delta.
  \forall f \in (\mathsf{arr}~A~B).
  \forall g \in (\mathsf{arr}~A~B).$ \\
  $(\forall x\in A. \mathsf{ap}~f~x = \mathsf{ap}~g~x) \rightarrow f = g.$
\item[${\mathbf{beta}}$:] $\forall A:\delta.\forall F:\iota\to\iota. \forall 
x\in A.(\mathsf{ap}~(\mathsf{lam}~A~F)~x) = F~x$.
\end{description}

If $\iota$ is interpreted using a model of ZFC and $\delta$ using a copy of 
the non-empty sets in this model, then the constants above can be interpreted 
in an obvious way so as to make the basic axioms true.

Given this theory, a basic translation from \holfour to \thfzero is
as follows.
Each \holfour type $\alpha$ (including type variables) is mapped to a term 
$\hat{\alpha}$ of type $\delta$.
\holfour type variables (constants) are mapped to \thfzero variables 
(constants) of type $\delta$.
For the remaining cases ${\mathsf{bool}}$, ${\mathsf{ind}}$, and $\mathsf{arr}$
are used.
Each \holfour term $s:\alpha$ is mapped to a \thfzero term $\hat{s}$ of 
type $\iota$, for which the context $\hat{s}\in\hat{\alpha}$ is always known.
The invariant can be maintained by including the hypothesis 
$\hat{x}\in\hat{\alpha}$ whenever $x$ is a variable or a constant. 
The $\mathsf{ap}$ and $\mathsf{lam}$ constants are used to handle \holfour 
applications and $\lambda$-abstractions.
The axioms ${\mathbf{ap_{tp}}}$ and ${\mathbf{lam_{tp}}}$ ensure the 
invariant is maintained.
Finally \holfour propositions (which may quantify over type variables) are 
translated to \thfzero propositions in an obvious way, using $\mathsf{p}$ 
to go from $\iota$ to $o$, and $\mathsf{i}_o$ to go from $o$ to $\iota$, when 
necessary.
As an added heuristic, the translation makes use of \thfzero connectives and 
quantifiers as deeply as possible, delaying the use of $\mathsf{p}$ whenever 
possible.

\vspace{-2mm}
\paragraph{Using Special Types.}
As with the first family of translations, the second family optimizes by
using special types for \holfour types with no type variables, e.g., 
${\mathtt{num}}$ and ${\mathtt{list}}~{\mathtt{num}}$.
Unlike the first family, special types are not used for monomorphic function 
types.
As a result it is not necessary to consider alternative $\mathsf{ap}$ 
operators.
A {\emph{basic monomorphic type}} is a monomorphic type that is not of 
the form $\alpha\to\beta$.
If special types are used, then for each basic monomorphic type occurring in 
a proposition a corresponding \thfzero type $\gamma$ is declared,
mappings and axioms relating $\gamma$ to the type $\iota$ of sets are declared,
and the type $\gamma$ is used to translate terms of the type and quantifiers 
over the type when possible.
For example, if a basic monomorphic type $\nu$ (e.g., ${\mathtt{num}}$) 
occurs in a \holfour proposition, then in addition to translating $\nu$ 
as a term $\hat{\nu}:\iota$ we also declare a \thfzero type $\tilde{\nu}$, 
${\mathsf{i}}_\nu:\tilde{\nu}\to\iota$ and 
${\mathsf{j}}_\nu:\iota\to\tilde\nu$ along with axioms
$\forall x:\tilde{\nu}.{\mathsf{j}}_\nu({\mathsf{i}}_\nu x) = x$
and
$\forall x:\iota.x\in \hat{\nu}\rightarrow {\mathsf{i}}_\nu({\mathsf{j}}_\nu x) 
= x$.

One obvious basic monomorphic type is $o$. 
In the case of $o$ a new type is not declared, but instead the \thfzero type 
{\tt \$o} is used. 
That is, $\tilde{o}$ denotes {\tt \$o}.
Note that $\mathsf{i}_o:\tilde{o}\to\iota$ is already declared.
Additionally, $\mathsf{j}_o$ is used as shorthand for $\mathsf{p}$,
which has the desired type $\iota\to\tilde{o}$.

Suppose a \holfour constant $c$ has type 
$\alpha_1\to\ldots \to\alpha_n\to \beta$, where 
$\alpha_1,\ldots,\alpha_n,\beta$ are basic monomorphic types with corresponding 
\thfzero types $\tilde{\alpha_1},\ldots,\tilde{\alpha_n},\tilde{\beta}$.
Instead of translating a term $c\,t_1\cdots t_n$ as a term of type 
$\iota$, each $t_i$ is translated to a term $\hat{t_i}$ of type 
$\tilde{\alpha_i}$, and a first order constant 
$\tilde{c}:\tilde{\alpha_1}\to\cdots\to\tilde{\alpha_n}\to\tilde{\beta}$ is 
used to translate to the term $\tilde{c}\,\hat{t_1}\cdots \hat{t_n}$ of 
type $\tilde{\beta}$.
In such a case an equation relating $\hat{c}$ to $\tilde{c}$ is also included.
Since the translation may return a term of type $\iota$ or $\tilde{\alpha}$, 
where $\alpha$ is a basic monomorphic type, $\mathsf{i}_\alpha$ and 
$\mathsf{j}_\alpha$ are used to obtain a term of type $\tilde{\alpha}$ or 
$\iota$ when one is required.
If a quantifier ranges over a monomorphic type $\alpha$, a quantifier
over type $\tilde{\alpha}$ is used instead of using a quantifier over 
type $\iota$ and using $\in$ to guard the quantifier.

\subsection{Translating to \tffzero}

There are two main modifications to the translation to \thfzero
when targeting \tffzero.
Firstly, propositions cannot be treated as special kinds of terms in \tffzero.
In order to deal with this $o$ is treated like other special types by
declaring a new type $\tilde{o}$ and functions 
${\mathsf{i}}_o:\tilde{o}\to\iota$ and ${\mathsf{j}}_o:\iota\to\tilde{o}$ 
along with corresponding axioms as above.
Note that unlike the \thfzero case, ${\mathsf{j}}_o$ differs from 
${\mathsf{p}}$.
In \tffzero ${\mathsf{p}}$ is a unary predicate on $\iota$, and 
${\mathsf{j}}_o$ is a function from $\iota$ to $\tilde{o}$.
In the \tffzero versions of the axioms ${\mathbf{Iso}}^1_o$ and 
${\mathbf{Iso}}^2_o$, ${\mathsf{p}}$ is replaced with ${\mathsf{j}}_o$.
Secondly, the background theory cannot include the higher-order 
${\mathsf{lam}}$ operator.
Therefore the ${\mathsf{lam}}$ operator is omitted, and lambda lifting is
used to translate (most) \holfour $\lambda$-abstractions.
The two higher-order axioms ${\mathbf{lam_{tp}}}$ and ${\mathbf{beta}}$
are also omitted.

In the \thfzero case, the background axioms are enough to infer the 
following (internal) propositional extensionality principle
$$\forall Q\in\mathsf{bool}. \forall R\in\mathsf{bool}. ({\mathsf{p}}~Q 
\leftrightarrow {\mathsf{p}}~R) \rightarrow Q = R$$
from the corresponding extensionality principle
$\forall Q R:o. (Q \leftrightarrow R) \rightarrow Q = R$
valid in \thfzero.
This is no longer the case in \tffzero, so propositional extensionality 
is added as an axiom.

There are two special cases where lambda lifting can be avoided:
identity and constant functions.
For this purpose a new unary function $\mathsf{I}$ on sets
and a new binary function $\mathsf{K}$ on sets are added.
Two new basic axioms are added to the ATP problem for these functions:
\begin{description}
\item[Id:] $\forall A:\delta.\forall X\in A.(\mathsf{ap}~(\mathsf{I}~A)~X) = X$.
\item[Const:] $\forall A:\delta.\forall Y:\iota.\forall X\in 
A.(\mathsf{ap}~(\mathsf{K}~A~Y)~X) = Y$.
\end{description}

A \holfour term $\lambda x:\alpha.x$ is translated as $\mathsf{I}~\hat{\alpha}$.
For a \holfour term $\lambda x:\alpha.t$, where $x$ is not free in $t$,
$t$ is translated to a first-order term $\hat{t}$ of type $\iota$,
and the $\lambda$-term is translated to $\mathsf{K}~\hat{\alpha}~\hat{t}$.
If there is already a function defined for $\lambda x:\alpha.t$ (with the 
same variable names), then that function is reused.
Otherwise, lambda lifting of $\lambda x:\alpha.t$ proceeds as follows.
Let $\alpha_1,\ldots,\alpha_m$ be type variables occurring in 
$\lambda x:\alpha.t$ and $y_1:\beta_1,\ldots,y_n:\beta_n$ be the free variables 
occurring in $\lambda x:\alpha.t$.
Assume $\hat{t}$ is a first-order term translation of $t$, with
$\hat{x}$ of type $\iota$ corresponding to the variable $x$.
(Note that this may have involved some lambda lifting.)
Let $\mathbf{f}$ be a new $m+n$-ary function returning sets.
If special types are not being used, then each argument of $\mathbf{f}$ is 
a set.
If special types are used, then each argument is a set unless it corresponds 
to $y_i:\beta_i$, where $\beta_i$ is a monomorphic type in which case the 
argument has type $\tilde{\beta_i}$.
The following axioms about $\mathbf{f}$ are added to the ATP problem:
\begin{description}
\item[$\mathbf{f}_{\mathbf{tp}}$:] $\forall A_1\cdots A_n:\delta.\forall 
Y_1\cdots Y_m:\iota.\cdots (\mathbf{f}~A_1\cdots~A_n~Y_1~\cdots~Y_m) \in 
\hat{\alpha}$.
\item[$\mathbf{f}_{\mathbf{beta}}$:] $\forall A_1\cdots A_n:\delta.\forall 
Y_1\cdots Y_m:\iota.\cdots\forall X\in\hat{\alpha}. 
\mathsf{ap}~(\mathbf{f}~A_1\cdots~A_n~Y_1~\cdots~Y_m)~X = \hat{t}$.
\end{description}
In these axioms the preconditions that each $Y_i$ must be in $\hat{\beta_i}$ 
if $Y_i$ has type $\iota$ have been elided (otherwise special types are being 
used, $\beta_i$ is monomorphic, $Y_i$ has type $\tilde{\beta_i}$, and no guard 
is required).

\subsection{Translating to \fof}

In order to translate to \fof, all terms must be translated to the same type, 
effectively the type $\iota$.
This requires omission of any special treatment of monomorphic types,
and instead all \holfour terms are translated to terms of type $\iota$.
The type $\delta$ of non-empty sets is also omitted.
Instead, $\iota$ is used wherever $\delta$ was used in the \tffzero setting,
and quantifiers that were over $\delta$ are guarded by a new non-emptiness 
predicate ${\mathsf{ne}}:\iota\to o$.
Aside from these changes, the translation proceeds using lambda lifting as 
in the \tffzero case.

\section{Case Study}
\label{case}

A very simple HOL4 theorem is 
$\forall f:\alpha\to\beta.\forall x:\alpha. {\mathsf{LET}}^{\alpha,\beta}~f~x = 
f~x$,
where ${\mathsf{LET}}^{\alpha,\beta}$ is defined to be 
$\lambda f:\alpha\to\beta.\lambda x:\alpha.f x$.
Informally the proof is clear: expand the definition of ${\mathsf{LET}}$ and 
perform two $\beta$-reductions. 
However, proving various translated versions of the problem range from 
trivial to challenging.

The first family of translations make use of a preprocessing step 
(Section~\ref{tg_tf1}) that 
changes the definition of ${\mathsf{LET}}$ from 
${\mathsf{LET}}^{\alpha,\beta} = \lambda f:\alpha\to\beta.\lambda x:\alpha.f x$
to
$$\forall x:\alpha\to\beta.\forall x':\alpha. 
{\mathsf{LET}}^{\alpha,\beta}~x~x' = x~x'.$$
This step makes the definition of ${\mathsf{LET}}$
the same (up to $\alpha$-conversion) as the theorem.
Even if further encodings are applied to obtain a first-order problem,
the axiom will still be the same as the conjecture.
Consequently all versions resulting from the first family of translations
are trivially provable.

The \thfzero-II version has conjecture
$$\forall A B:\delta.\forall f\in ({\mathsf{arr}}~A~B).\forall x\in 
A.{\mathsf{ap}}~({\mathsf{ap}}~({\mathsf{LET}}~A~B)~f)~x = {\mathsf{ap}}~f~x$$
and the axiom (corresponding to the definition of ${\mathsf{LET}}$)
$$\forall A B:\delta.{\mathsf{LET}}~A~B = 
{\mathsf{lam}}~({\mathsf{arr}}~A~B)~(\lambda f:\iota.{\mathsf{lam}}~A~(\lambda 
x:\iota.{\mathsf{ap}}~f~x)).$$
The axiom defining ${\mathsf{LET}}$ combined with the basic axiom 
${\mathbf{beta}}$ is enough to prove the theorem. 
However, the \thfzero-II version also includes all the other basic axioms along 
with internal versions of the logical constants for universal quantification 
and equality.
The extra axioms make the problem hard for ATP systems, but if only the
necessary axioms are provided the problem is easy.
In \tffzero-II and \fof-II the conjecture is the same as in the \thfzero-II 
version, but the definition of ${\mathsf{LET}}$ is split into two functions 
declared when lambda lifting:
$$\forall A\cdots\forall B\cdots{\mathsf{LET}}~A~B = {\mathbf{f_{14}}}~A~B,$$
$$\forall A\cdots\forall B\cdots\forall f\in ({\mathsf{arr}}~A~B). 
{\mathsf{ap}}~({\mathbf{f_{14}}}~A~B)~f = {\mathbf{f_{13}}}~A~B~f$$
and
$$\forall A\cdots\forall B\cdots\forall f\in ({\mathsf{arr}}~A~B). \forall x\in 
A. {\mathsf{ap}}~({\mathbf{f_{13}}}~A~B~f)~x = {\mathbf{ap}}~f~x.$$
All the first-order versions of this problem are easy for current ATP systems.

\section{Results}
\label{results}

Since the HOL4 library has a natural order of 
the problems, each translation can generate two versions of each problem. 
The \emph{bushy} (small) version contains only the (translated) library 
facts that were needed for the HOL4 proof of the theorem.
The \emph{chainy} (large) version contains all the facts that precede the
theorem in the library order, i.e., the real task faced by hammer systems. 
Chainy problems typically include thousands of axioms, requiring the use 
of \emph{premise selection} algorithms~\cite{abs-1108-3446} as a front-end in
the ATP systems. 
Thus, in order to maintain the focus on ATP system performance, the results
of running the ATP systems on the bushy problems are presented here.

Nineteen ATPs were run on the 12140 problems in each of the
bushy problem sets, according to the ATPs' support for the various TPTP 
formats. In each case we ran the ATP with a CPU time limit of 60s per problem.
Table~\ref{tab:results-by-format} summarizes the results.
In union, more proofs were found in the first family of translations than 
in the second family, in all formats.
However, some provers like Vampire 4.3
and SPASS 3.9 do better on \fof-II than on \fof-I. 
This indicates that these provers are probably better at reasoning with type 
guards than with type tags.
Of the 12140 problems 7412 (61.1\%) were solved by some ATP in one of the 
representations.

\begin{table*}[h]
\begin{scriptsize}
\begin{center}
\ra{1.3}
\begin{tabular}{l|cccccccc|c}
 System & \thfone-I & \thfzero-I & \thfzero-II & \tffone-I & \tffzero-I & 
 \tffzero-II & \fof-I & \fof-II & {\bf Union} \\
\midrule
agsyHOL 1.0~\cite{DBLP:conf/cade/Lindblad14}    &           & 1374       & 
1187        &           &            &             &        &         & 1605\\
Beagle 0.9.47~\cite{dblp:conf/cade/baumgartnerw13}     &           &            
&             &           & 2008       & 2047        & 2449   & 2498    & 3183\\
cocATP 0.2.0 &           & 899        & 599         &           &            
&             &        &         & 1000\\
CSE\_E 1.0~\cite{XL+18}     &           &            &             &           
&            &             & 4251   & 3102    & 4480\\
CVC4 1.6~\cite{barrett:2011:cvc:2032305.2032319}       &           &            
&             &           & 4851       & 3991        & 5030   & 3746    & 5709\\
E 2.2~\cite{Schulz13}          &           &            &             
&           & 4277       & 3622        & 4618   & 3844    & 5118\\
HOLyHammer 0.21~\cite{holyhammer} & 5059      &            &             
&           &            &             &        &         & 5059\\
iProver 2.8~\cite{Korovin08}    &           &            &             
&           &            &             & 2778   & 2894    & 3355\\
iProverModulo 2.5-0.1~\cite{burel11experimenting} &           &            
&             &           & 2435       & 1639        & 1433   & 1263    & 2852\\
LEO-II 1.7.0~\cite{BP+08}     &           & 2579       & 1923        
&           &            &             & 2119   & 1968    & 3702\\
Leo-III 1.3~\cite{DBLP:conf/lpar/SteenWB17,SteenBenz2018}    & 6668      & 
5018       & 3485        & 3458      & 4032       & 3421        & 3986   & 
3185    & 7090\\
Metis 2.4~\cite{hurd05}      &           &            &             &           
&            &             & 2353   & 474     & 2356\\
Princess 170717~\cite{Rue08,Rue12}   &           &            &             
&           & 3646       & 2138        & 3162   & 2086    & 4096\\
Prover9 1109a~\cite{McC-Prover9-URL}    &           &            &             
&           &            &             & 2894   & 1742    & 3128\\
Satallax 3.3~\cite{Brown12}   &           & 2207       & 1292        
&           &            &             &        &         & 2494\\
SPASS 3.9~\cite{WeidenbachDFKSW09}      &           &            &             
&           &            &             & 2850   & 3349    & 3821\\
Vampire 4.3~\cite{Vampire}    &           &            &             
&           & 4837       & 4693        & 4008   & 4928    & 5929\\
ZenonModulo~\cite{DBLP:conf/lpar/DelahayeDGHH13} &           &            
&             & 1071      & 1038       & 1041        & 1026   & 1198    & 1751\\
Zipperposition 1.4~\cite{DBLP:phd/hal/Cruanes15} &           & 2252       & 
2161        & 3771      & 3099       & 2576        & 2531   & 1795    & 4251\\
\bottomrule
{\bf Union}&   6824    & 5209       & 3771        & 4663      & 5732       & 
5074        & 5909   & 5249    & {\bf 7412}\\
\end{tabular}
\end{center}
\caption{\label{tab:results-by-format} Number of theorems proved, out of 12140.
Each ATP is evaluated on all its supported TPTP formats.}
\end{scriptsize}
\vspace{-9mm}
\end{table*}

The \tactictoe~\cite{GauthierKU17,gkukn-toappear-jar18} prover built into 
\holfour has been tested as a baseline comparison, and it (re)proves
5327 of 8855 chainy versions of the problems (60.2\%).
\tactictoe is a machine-learning guided prover that searches for a tactical 
proof by selecting suitable tactics and theorems learned from human-written 
tactical proofs.
By design, this system works in the chainy setting.
In total 8840 (72.8\%) of the 12140 problems can be proved by either \tactictoe
or one of the ATPs using one of the translations.

\section{GRUNGE as CASC LTB Division}
\label{CASCLTB}

The CADE ATP System Competition (CASC) \cite{Sut16} is the annual evaluation 
of fully automatic, classical logic Automated Theorem Proving (ATP) systems --
the world championship for such systems.
CASC is divided into divisions according to problem and system characteristics.
Each competition division uses problems that have certain logical,
language, and syntactic characteristics, so that the systems that
compete in the division are, in principle, able to attempt all the
problems in the division.
For example, the First-Order Form (\fof) division uses problems in full
first-order logic, with each problem having axioms and a conjecture to
be proved.

While most of the CASC divisions present the problems to the ATP systems one
at a time, with an individual CPU or wall clock time limit per problem,
the Large Theory Batch (LTB) division presents the problems in batches, 
with an overall wall clock time limit on the batch.
As the name also suggests, the problems in each batch come from a ``large 
theory'', which typically has many functors and predicates, and many axioms 
of which only a few are required for the proof of a theorem.
The problems in a batch typically have a common core set of axioms used
by all problems, and each problem typically has additional axioms that 
are specific to the problem.
The batch presentation allows the ATP systems to load and preprocess the
common core set of axioms just once, and to share logical and control
results between proof searches.
Each batch is accompanied by a set of training problems and their solutions, 
taken from the same source as the competition problems. 
The training data can be used for ATP system tuning and learning during 
(typically at the start of) the competition.

In CASC-J9 \cite{Sutcliffe18} -- the most recent edition of the competition -- 
the LTB division used \fof problems exported from \cakeml~\cite{KumarMNO14}.
At the time there was growing interest in an LTB division for typed
higher-order problems, and it became evident that a multi-format LTB 
division would add a valuable dimension to CASC.
For the CASC-27 LTB division each problem was presented in multiple 
formats: \thfone, \thfzero, \tffone, \tffzero, and \fof. 
The work described in this paper provides the problems.
Systems were able to attempt whichever versions they support, and a 
solution to any version constitutes a solution to the problem.
For example, Leo-III is able to handle all the formats, while 
E can attempt only the \\tffzero and \fof formats.

The batch presentation of problems in the LTB division provides interesting 
opportunities for ATP systems, including making multiple attempts on problems
and learning search heuristics from proofs found.
The multi-format LTB division extends these possibilities by allowing
multiple attempts on problems by virtue of the multiple formats
available, and learning from proofs found in one format to improve
performance on problems in another format.
The latter is especially interesting, with little known research in
this direction.

\section{Related Work}\label{relatedwork}

The \holfour library already has translations for SMT solvers such as 
Yices~\cite{WeberTjark}, Z3~\cite{Z3reconstruction} and Beagle. 
A link to first-order ATPs is also available thanks to exports \cite{hh4h4}
of \holfour theories to the \holyhammer framework~\cite{holyhammer}.
Another notable project that facilitates the export of \holfour
theories is \opentheory~\cite{opentheory}.
The general approach for higher-order to first-order translations is laid 
out in Hurd~\cite{hurd03}.
An evaluation of the effect of different  translations on ATP-provability was 
performed in~\cite{MengP08}. 
A further study shows the potential improvements provided by the use of 
supercombinators~\cite{DBLP:conf/cpp/Czajka16}.
In our work, the use of lambda-lifting (or combinators) is not necessary 
in \thfzero-II thanks to the use of the higher-order operator $\mathsf{lam}$.
This is similar to using higher-order abstract syntax to model syntax with 
binders~\cite{Pfenning1988}.

A method for encoding of polymorphic types as terms through type tags (as 
in our first translation) or type guards (as in our second translation) is 
described in \cite{blanchettebps13}.
Translations~\cite{meson,DBLP:conf/lpar/SteenWB17} from a polymorphic logic 
to a monomorphic poly-sorted logic without encoding typically rely on 
heuristic instantiations of type variables. 
However, heuristics may miss useful instantiations, and make the translation 
less modular (i.e., context dependent).
Our translations to \thfzero and \tffzero try to get the best of both worlds 
by using a type encoding for polymorphic types and special types for basic 
monomorphic types.

\section{Conclusion}

This work has defined, compared, and evaluated ATP performance on two families 
of translations of the \holfour logic to a number of ATP formalisms, and 
described a new unified large-theory ATP benchmark (GRUNGE) based on them.  
The first family is designed to play to the strengths of the calculi of 
most ATP systems, while the second family is based on more straightforward 
semantics rooted in set theory.
The case study shows how different the translated problems may be,
even in a simple example.
A number of methods and optimizations have been used, however it
is clear that the translations can be further optimized and that
different encodings favour different provers.
Out of 12140 \holfour theorems, the ATP systems can solve 
7412 problems in one or more of the formats.
The \tactictoe system that works directly in the \holfour formalism and 
uses \holfour tactics could solve 5327 problems. 
Together the total number of problems solved is 8840.
Leo-III was the strongest system in the higher-order representations.
In the first-order representations the strongest systems
were Zipperposition, CVC4, E and Vampire.
A pre-release of the bushy versions of the problems
was provided before CASC-27\footnote{%
\url{http://www.tptp.org/CASC/27/TrainingData.HL4.tgz}}, to 
allow system developers to adapt and tune their systems before the 
competition.


\begin{thebibliography}{10}
\providecommand{\url}[1]{\texttt{#1}}
\providecommand{\urlprefix}{URL }
\providecommand{\doi}[1]{https://doi.org/#1}

\bibitem{abs-1108-3446}
Alama, J., Heskes, T., K\"{u}hlwein, D., Tsivtsivadze, E., Urban, J.: Premise
  selection for mathematics by corpus analysis and kernel methods. J. Autom.
  Reasoning  \textbf{52}(2),  191--213 (2014). \doi{10.1007/s10817-013-9286-5}

\bibitem{barrett:2011:cvc:2032305.2032319}
Barrett, C., Conway, C.L., Deters, M., Hadarean, L., Jovanovi\'{c}, D., King,
  T., Reynolds, A., Tinelli, C.: {CVC}4. In: Gopalakrishnan, G., Qadeer, S.
  (eds.) Conference on Computer Aided Verification (CAV). Lecture Notes in
  Computer Science, vol.~6806, pp. 171--177. Springer (2011),
  \url{https://doi.org/10.1007/978-3-642-22110-1_14}

\bibitem{dblp:conf/cade/baumgartnerw13}
Baumgartner, P., Waldmann, U.: Hierarchic superposition with weak abstraction.
  In: Bonacina, M.P. (ed.) Conference on Automated Deduction (CADE). Lecture
  Notes in Computer Science, vol.~7898, pp. 39--57. Springer (2013),
  \url{http://dx.doi.org/10.1007/978-3-642-38574-2_3}

\bibitem{BP+08}
Benzm{\"u}ller, C., Paulson, L., Theiss, F., Fietzke, A.: {LEO-II -- A
  Cooperative Automatic Theorem Prover for Higher-Order Logic}. In:
  Baumgartner, P., Armando, A., Dowek, G. (eds.) {Proceedings of the 4th
  International Joint Conference on Automated Reasoning}. pp. 162--170.
  No.~5195 in Lecture Notes in Artificial Intelligence, Springer-Verlag (2008)

\bibitem{C25}
Benzm{\"u}ller, C., Rabe, F., Sutcliffe, G.: {THF0} -- the core of the {TPTP}
  language for classical higher-order logic. In: Armando, A., Baumgartner, P.,
  Dowek, G. (eds.) Automated Reasoning, 4th International Joint Conference,
  IJCAR 2008, Sydney, Australia, August 12-15, 2008, Proceedings. Lecture Notes
  in Computer Science, vol.~5195, pp. 491--506. Springer (2008),
  \url{http://christoph-benzmueller.de/papers/C25.pdf}

\bibitem{blanchettebps13}
Blanchette, J.C., B\"ohme, S., Popescu, A., Smallbone, N.: Encoding monomorphic
  and polymorphic types. In: Piterman, N., Smolka, S.A. (eds.) Conference on
  Tools and Algorithms for the Construction and Analysis of Systems (TACAS).
  Lecture Notes in Computer Science, vol.~7795, pp. 493--507. Springer (2013),
  \url{https://doi.org/10.1007/978-3-642-36742-7_34}

\bibitem{hammers4qed}
Blanchette, J.C., Kaliszyk, C., Paulson, L.C., Urban, J.: Hammering towards
  {QED}. J. Formalized Reasoning  \textbf{9}(1),  101--148 (2016).
  \doi{10.6092/issn.1972-5787/4593}

\bibitem{tff1}
Blanchette, J.C., Paskevich, A.: {TFF1: The TPTP} typed first-order form with
  rank-1 polymorphism. In: Bonacina, M.P. (ed.) CADE. Lecture Notes in Computer
  Science, vol.~7898, pp. 414--420. Springer (2013)

\bibitem{Z3reconstruction}
B\"ohme, S., Weber, T.: Fast {LCF}-style proof reconstruction for {Z3}. In:
  Kaufmann, M., Paulson, L.C. (eds.) Conference on Interactive Theorem Proving
  (ITP). Lecture Notes in Computer Science, vol.~6172, pp. 179--194. Springer
  (2010), \url{http://dx.doi.org/10.1007/978-3-642-14052-5_14}

\bibitem{Brown12}
Brown, C.E.: Satallax: An automatic higher-order prover. In: Gramlich, B.,
  Miller, D., Sattler, U. (eds.) IJCAR. Lecture Notes in Computer Science,
  vol.~7364, pp. 111--117. Springer (2012)

\bibitem{burel11experimenting}
Burel, G.: Experimenting with deduction modulo. In: Sofronie-Stokkermans, V.,
  Bjørner, N. (eds.) CADE 2011. Lecture Notes in Artificial Intelligence,
  vol.~6803, pp. 162--176. Springer (2011)

\bibitem{Church:SimplyTyped}
Church, A.: A formulation of the simple theory of types. J. Symbolic Logic
  \textbf{5},  56--68 (1940)

\bibitem{DBLP:phd/hal/Cruanes15}
Cruanes, S.: Extending Superposition with Integer Arithmetic, Structural
  Induction, and Beyond. (Extensions de la Superposition pour
  l'Arithm{\'{e}}tique Lin{\'{e}}aire Enti{\`{e}}re, l'Induction Structurelle,
  et bien plus encore). Ph.D. thesis, {\'{E}}cole Polytechnique, Palaiseau,
  France (2015), \url{https://tel.archives-ouvertes.fr/tel-01223502}

\bibitem{DBLP:conf/cpp/Czajka16}
Czajka, L.: Improving automation in interactive theorem provers by efficient
  encoding of lambda-abstractions. In: Avigad, J., Chlipala, A. (eds.)
  Proceedings of the 5th {ACM} {SIGPLAN} Conference on Certified Programs and
  Proofs, Saint Petersburg, FL, USA, January 20-22, 2016. pp. 49--57. {ACM}
  (2016). \doi{10.1145/2854065.2854069}

\bibitem{DBLP:conf/lpar/DelahayeDGHH13}
Delahaye, D., Doligez, D., Gilbert, F., Halmagrand, P., Hermant, O.: Zenon
  modulo: When achilles outruns the tortoise using deduction modulo. In:
  McMillan et~al.  \cite{DBLP:conf/lpar/2013}, pp. 274--290.
  \doi{10.1007/978-3-642-45221-5\_20},
  \url{https://doi.org/10.1007/978-3-642-45221-5\_20}

\bibitem{hh4h4}
Gauthier, T., Kaliszyk, C.: Premise selection and external provers for {HOL4}.
  In: Certified Programs and Proofs (CPP'15). Lecture Notes in Computer
  Science, Springer (2015). \doi{10.1145/2676724.2693173}

\bibitem{GauthierKU17}
Gauthier, T., Kaliszyk, C., Urban, J.: {TacticToe}: Learning to reason with
  {HOL4} tactics. In: Eiter, T., Sands, D. (eds.) LPAR-21, 21st International
  Conference on Logic for Programming, Artificial Intelligence and Reasoning,
  Maun, Botswana, May 7-12, 2017. EPiC Series in Computing, vol.~46, pp.
  125--143. EasyChair (2017),
  \url{http://www.easychair.org/publications/paper/340355}

\bibitem{gkukn-toappear-jar18}
Gauthier, T., Kaliszyk, C., Urban, J., Kumar, R., Norrish, M.: Learning to
  prove with tactics. CoRR  (2018), \url{http://arxiv.org/abs/1804.00596}

\bibitem{Gordon:1993:ITH}
Gordon, M.J.C., Melham, T.F. (eds.): Introduction to {HOL}: A theorem proving
  environment for higher order logic. Cambridge University Press (1993),
  \url{http://www.cs.ox.ac.uk/tom.melham/pub/Gordon-1993-ITH.html}

\bibitem{Harrison96}
Harrison, J.: {HOL Light}: A tutorial introduction. In: Srivas, M.K.,
  Camilleri, A.J. (eds.) FMCAD. Lecture Notes in Computer Science, vol.~1166,
  pp. 265--269. Springer (1996)

\bibitem{meson}
Harrison, J.: Optimizing proof search in model elimination. In: McRobbie, M.,
  Slaney, J. (eds.) Conference on Automated Deduction (CADE). pp. 313--327.
  No.~1104 in LNAI, Springer (1996),
  \url{https://doi.org/10.1007/3-540-61511-3_97}

\bibitem{HarrisonUW14}
Harrison, J., Urban, J., Wiedijk, F.: History of interactive theorem proving.
  In: Siekmann, J.H. (ed.) Computational Logic, Handbook of the History of
  Logic, vol.~9, pp. 135--214. Elsevier (2014).
  \doi{10.1016/B978-0-444-51624-4.50004-6}

\bibitem{hurd03}
Hurd, J.: First-order proof tactics in higher-order logic theorem provers.
  Design and Application of Strategies/Tactics in Higher Order Logics, number
  NASA/CP-2003-212448 in NASA Technical Reports pp. 56--68 (2003)

\bibitem{hurd05}
Hurd, J.: System description: {T}he {M}etis proof tactic. In:
  Christoph~Benzmueller, John~Harrison, C.S. (ed.) Workshop on Empirically
  Successful Automated Reasoning in Higher-Order Logic (ESHOL). pp. 103--104
  (2005), \url{https://arxiv.org/pdf/cs/0601042}

\bibitem{opentheory}
Hurd, J.: The {OpenTheory} standard theory library. In: Bobaru, M.G., Havelund,
  K., Holzmann, G.J., Joshi, R. (eds.) NASA Formal Methods. Lecture Notes in
  Computer Science, vol.~6617, pp. 177--191. Springer (2011),
  \url{http://dx.doi.org/10.1007/978-3-642-20398-5_14}

\bibitem{KSR16}
Kaliszyk, C., Sutcliffe, G., Rabe, F.: {TH1: The TPTP Typed Higher-Order Form
  with Rank-1 Polymorphism}. In: Fontaine, P., Schulz, S., Urban, J. (eds.)
  {Proceedings of the 5th Workshop on Practical Aspects of Automated
  Reasoning}. pp. 41--55. No.~1635 in CEUR Workshop Proceedings (2016)

\bibitem{holyhammer}
Kaliszyk, C., Urban, J.: Learning-assisted automated reasoning with {F}lyspeck.
  J. Autom. Reasoning  \textbf{53}(2),  173--213 (2014).
  \doi{10.1007/s10817-014-9303-3}

\bibitem{king1996development}
King, D., Arthan, R., Winnersh, I.: Development of practical verification
  tools. ICL Systems Journal  \textbf{11},  106--122 (1996)

\bibitem{Korovin08}
Korovin, K.: {iProver} -- an instantiation-based theorem prover for first-order
  logic (system description). In: Armando, A., Baumgartner, P., Dowek, G.
  (eds.) Automated Reasoning, 4th International Joint Conference, {IJCAR} 2008,
  Sydney, Australia, August 12-15, 2008, Proceedings. Lecture Notes in Computer
  Science, vol.~5195, pp. 292--298. Springer (2008)

\bibitem{Vampire}
Kov{\'a}cs, L., Voronkov, A.: First-order theorem proving and {V}ampire. In:
  Sharygina, N., Veith, H. (eds.) CAV. Lecture Notes in Computer Science,
  vol.~8044, pp. 1--35. Springer (2013)

\bibitem{KumarMNO14}
Kumar, R., Myreen, M.O., Norrish, M., Owens, S.: {CakeML}: a verified
  implementation of {ML}. In: Jagannathan, S., Sewell, P. (eds.) The 41st
  Annual {ACM} {SIGPLAN-SIGACT} Symposium on Principles of Programming
  Languages, {POPL}'14, San Diego, CA, USA, January 20-21, 2014. pp. 179--192.
  {ACM} (2014). \doi{10.1145/2535838.2535841}

\bibitem{DBLP:conf/cade/Lindblad14}
Lindblad, F.: A focused sequent calculus for higher-order logic. In: Demri, S.,
  Kapur, D., Weidenbach, C. (eds.) Automated Reasoning -- 7th International
  Joint Conference, {IJCAR} 2014, Held as Part of the Vienna Summer of Logic,
  {VSL} 2014, Vienna, Austria, July 19-22, 2014. Proceedings. Lecture Notes in
  Computer Science, vol.~8562, pp. 61--75. Springer (2014).
  \doi{10.1007/978-3-319-08587-6\_5},
  \url{https://doi.org/10.1007/978-3-319-08587-6\_5}

\bibitem{McC-Prover9-URL}
McCune, W.: {Prover9 and Mace4} (2005--2010),
  \url{http://www.cs.unm.edu/~mccune/prover9/}

\bibitem{DBLP:conf/lpar/2013}
McMillan, K.L., Middeldorp, A., Voronkov, A. (eds.): Logic for Programming,
  Artificial Intelligence, and Reasoning -- 19th International Conference,
  LPAR-19, Stellenbosch, South Africa, December 14-19, 2013. Proceedings,
  Lecture Notes in Computer Science, vol.~8312. Springer (2013)

\bibitem{MengP08}
Meng, J., Paulson, L.C.: Translating higher-order clauses to first-order
  clauses. J. Autom. Reasoning  \textbf{40}(1),  35--60 (2008)

\bibitem{NPW02}
Nipkow, T., Paulson, L., Wenzel, M.: {Isabelle/HOL: A Proof Assistant for
  Higher-Order Logic}. No.~2283 in Lecture Notes in Computer Science,
  Springer-Verlag (2002)

\bibitem{Pfenning1988}
Pfenning, F., Elliot, C.: Higher-order abstract syntax. In: Proceedings of the
  ACM SIGPLAN 1988 Conference on Programming Language Design and
  Implementation. pp. 199--208. PLDI '88, ACM, New York, NY, USA (1988).
  \doi{10.1145/53990.54010}

\bibitem{Pitts93}
Pitts, A.: The {HOL} logic. In: Gordon and Melham  \cite{Gordon:1993:ITH},
  \url{http://www.cs.ox.ac.uk/tom.melham/pub/Gordon-1993-ITH.html}

\bibitem{Rue08}
R{\"u}mmer, P.: {A Constraint Sequent Calculus for First-Order Logic with
  Linear Integer Arithmetic}. In: Cervesato, I., Veith, H., Voronkov, A. (eds.)
  {Proceedings of the 15th International Conference on Logic for Programming,
  Artificial Intelligence, and Reasoning}. pp. 274--289. No.~5330 in Lecture
  Notes in Artificial Intelligence, Springer-Verlag (2008)

\bibitem{Rue12}
R{\"u}mmer, P.: {E-Matching with Free Variables}. In: Bj{\o}rner, N., Voronkov,
  A. (eds.) {Proceedings of the 18th International Conference on Logic for
  Programming, Artificial Intelligence, and Reasoning}. pp. 359--374. No.~7180
  in Lecture Notes in Artificial Intelligence, Springer-Verlag (2012)

\bibitem{Schulz13}
Schulz, S.: System description: {E} 1.8. In: McMillan et~al.
  \cite{DBLP:conf/lpar/2013}, pp. 735--743. \doi{10.1007/978-3-642-45221-5\_49}

\bibitem{SlindN08}
Slind, K., Norrish, M.: A brief overview of {HOL4}. In: Mohamed, O.A.,
  Mu{\~n}oz, C.A., Tahar, S. (eds.) Theorem Proving in Higher Order Logics,
  21st International Conference, TPHOLs 2008, Montreal, Canada, August 18-21,
  2008. Proceedings. Lecture Notes in Computer Science, vol.~5170, pp. 28--32.
  Springer (2008)

\bibitem{SteenBenz2018}
Steen, A., Benzm{\"u}ller, C.: The higher-order prover {Leo-III}. In: Galmiche,
  D., Schulz, S., Sebastiani, R. (eds.) Automated Reasoning. IJCAR 2018.
  Lecture Notes in Computer Science, vol. 10900, pp. 108--116. Springer, Cham
  (2018), \url{http://christoph-benzmueller.de/papers/C70.pdf}

\bibitem{DBLP:conf/lpar/SteenWB17}
Steen, A., Wisniewski, M., Benzm{\"{u}}ller, C.: Going polymorphic -- {TH1}
  reasoning for {Leo-III}. In: Eiter, T., Sands, D., Sutcliffe, G., Voronkov,
  A. (eds.) IWIL@LPAR 2017 Workshop and {LPAR-21} Short Presentations, Maun,
  Botswana, May 7-12, 2017. Kalpa Publications in Computing, vol.~1. EasyChair
  (2017), \url{http://www.easychair.org/publications/paper/346851}

\bibitem{Sut16}
Sutcliffe, G.: {The CADE ATP System Competition -- CASC}. AI Magazine
  \textbf{37}(2),  99--101 (2016)

\bibitem{Sutcliffe17}
Sutcliffe, G.: {The TPTP Problem Library and Associated Infrastructure. From
  CNF to TH0, TPTP v6.4.0}. Journal of Automated Reasoning  \textbf{59}(4),
  483--502 (2017)

\bibitem{SS+12}
Sutcliffe, G., Schulz, S., Claessen, K., Baumgartner, P.: {The TPTP Typed
  First-order Form with Arithmetic}. In: Bj{\o}rner, N., Voronkov, A. (eds.)
  {Proceedings of the 18th International Conference on Logic for Programming,
  Artificial Intelligence, and Reasoning}. pp. 406--419. No.~7180 in Lecture
  Notes in Artificial Intelligence, Springer-Verlag (2012)

\bibitem{Sutcliffe18}
Sutcliffe, G.: The 9th {IJCAR} automated theorem proving system competition --
  {CASC-J9}. {AI} Commun.  \textbf{31}(6),  495--507 (2018).
  \doi{10.3233/AIC-180773}

\bibitem{WeberTjark}
Weber, T.: {SMT} solvers: new oracles for the {HOL} theorem prover.
  International Journal on Software Tools for Technology Transfer
  \textbf{13}(5),  419--429 (2011),
  \url{http://dx.doi.org/10.1007/s10009-011-0188-8}

\bibitem{WeidenbachDFKSW09}
Weidenbach, C., Dimova, D., Fietzke, A., Kumar, R., Suda, M., Wischnewski, P.:
  {SPASS Version 3.5}. In: Schmidt, R.A. (ed.) CADE. Lecture Notes in Computer
  Science, vol.~5663, pp. 140--145. Springer (2009)

\bibitem{XL+18}
Xu, Y., Liu, J.~Chen, S., Zhong, X., He, X.: {Contradiction Separation Based
  Dynamic Multi-clause Synergized Automated Deduction}. Information Sciences
  \textbf{462},  93--113 (2018)

\end{thebibliography}
\end{document}